\documentstyle[fleqn,epsfig]{aipproc}
\newcommand{\ds}{\displaystyle}
\newcommand{\st}{\scriptstyle}

\newcommand{\cl}{\centering}
\newcommand{\nn}{\nonumber}

\newcommand{\pslash}{p\hspace{-.5em}/\hspace{.15em}}  
\title{Form Factors of Baryons \\
in a Confining and Covariant Diquark-Quark Model\footnote{
Talk given by M.O. at the Workshop on ''Effective Theories of Low
Energy QCD'', Coimbra, Portugal, Sep 10-15 1999.
Supported by the 
BMBF (06--TU--888)
and by the DFG (We 1254/4-1).} }
\author{M.~Oettel, S.~Ahlig, R.~Alkofer,
and C.~Fischer}
\address{Institute for Theoretical Physics, T{\"u}bingen University, \\
Auf der Morgenstelle 14, D-72076 T{\"u}bingen}
\begin{document}
\rightline{UNITUE--THEP--15/99}
\rightline{nucl-th/9910079}
\maketitle \noindent
\begin{abstract}
We treat baryons as bound states of scalar or axialvector diquarks and
a constituent quark which interact through quark exchange.
This description results as an approximation to the relativistic
Faddeev equation for three quarks which yields an effective Bethe-Salpeter
equation.
Octet and decuplet masses and fully four-dimensional wave functions
have been computed for two cases: assuming an essentially pointlike diquark
 on the one
hand, and a diquark with internal structure on the other hand.
Whereas the differences in the mass spectrum are fairly small, the nucleon
electromagnetic form factors are greatly improved assuming a diquark with
structure. First calculations to the pion-nucleon form factor
also suggest improvements. 
\end{abstract}
\section{Motivation}
Two approaches to the rich structure of strong interaction phenomena
have been the topic of this workshop. The first one, effective theories
like Chiral Perturbation Theory, resorts to including only physical fields
with a suitable expansion parameter. The second approach, the building
of effective models, often tries to interpolate between QCD and
observable degrees of freedom by taking loans from the latter 
in terms of the assumed relevant degrees of freedom, such as 
(constituent) quarks.
Different types of these models describe
various aspects of baryon physics. Among them are
nonrelativistic quark models, various sorts of bag models and approaches
describing baryons by means of collective variables like topological or non-topological
solitons \cite{Bha88}. Most of these models
are designed to work in the low-energy region and generally do not match the calculations within perturbative 
QCD. Considering the great
experimental progress in the medium energy range
with momentum transfers between 1 and 5 GeV$^2$, 
there is a high demand for models describing 
baryon physics in this region that connects the low and high energy regimes. 

To describe this kind of physics,
a fully covariant approach seems
indispensable. 
Furthermore, the effects of quark confinement should be incorporated
into a reliable description to avoid unphysical break-ups of baryons
into their constituents. This is in sharp contrast to 
low-energy or static observables: baryon masses and magnetic moments,
{\em e.g.}, can be understood in terms of a dynamically generated
constituent quark mass through chiral symmetry breaking. 
Confinement plays seemingly an unimportant role.
 
The Nambu-Jona-Lasinio model in its various guises shows this feature
of a dynamically generated quark mass
and has thus been utilized to describe mesonic properties quite 
successfully \cite{Ebe94}.
The description of baryons within this model allows for two
possibilities: They may appear as non-topological solitons
 \cite{Alk96,Chr96} or as bound states of quark and diquark \cite{Rei90}. 
In ladder approximation, diquarks appear as poles in quark-quark scattering
and therefore as physical particles. They are confined when going
beyond ladder approximation \cite{Hel97b}.
A study which incorporates both, solitons and diquark-quark bound states
 \cite{Zuc97}, shows
that the mesonic cloud and the quark-diquark interaction
contribute about equally to the binding energy of the baryon.

On the other hand, the relativistic three-body problem
can be simplified when discarding three-body irreducible 
interactions. The resulting Faddeev-type problem can be reduced further 
by assuming separable two-quark correlations which are usually called diquarks
\cite{Bur89,Ish95}. The Faddeev equations then collapse
to a Bethe-Salpeter equation whose solutions describe the baryons.
Quark and diquark hereby interact through quark exchange which restores
full antisymmetry between the three 
quarks\footnote{Due to antisymmetry in the color indices and the related 
symmetrization of all other quantum numbers the Pauli principle
leads to an attractive interaction in contrast to ''Pauli
repulsion'' known in conventional few-fermion systems.}. It is 
interesting to note that within the
NJL model the two-quark correlations 
(or 4-point quark Green function) are separable in first order
to yield a sum over poles  of diquarks with different quantum numbers.
 In analogy
to the meson spectrum\footnote{Scalar diquarks correspond to 
pseudoscalar mesons and axialvector diquarks to vector mesons
due to the intrinsically different parity of a fermion-antifermion pair
compaired to a fermion pair.},
scalar and axialvector diquarks are
assumed to be the lowest-lying and
thus the most important particles. This line of approach has been taken in
\cite{Ish95}.

\section{The Model}

In the subsequent sections, we will follow this approach and derive
an effective baryon Bethe-Salpeter equation with quark and diquark
as constituents.
However, to mimic confinement,
we will avoid the diquark poles which would correspond
to unphysical thresholds. To this end,
consider the 4-point quark Green function in coordinate space,
\begin{equation}
G_{\alpha\beta\gamma\delta}(x_1,x_2,x_3,x_4) =   
 \langle
T(q_\gamma(x_3) q_\alpha(x_1) \bar q_\beta(x_2) \bar q_\delta(x_4)) \rangle
\; , \label{Gen4q-G}
\end{equation}
where $\alpha, \beta, \gamma$, and $\delta$ denote the Dirac indices of
the quarks.
Assuming this 4-point function to be separable, we will parameterize
scalar and axialvector diquark correlations as:
\begin{eqnarray}
G_{\alpha\gamma , \beta\delta}^{\hbox{\tiny sep}}(p,q,P) \, &:=&
 e^{-iPY}  \, \int d^4\!X\, d^4\!y\, d^4\!z \,\,  e^{iqz}
e^{-ipy} e^{iPX} 
 G_{\alpha\beta\gamma\delta}^{\hbox{\tiny sep}} (x_1,x_2,x_3,x_4) 
 \label{dq_sep}
 \\
&=& 
 \chi_{\gamma\alpha}(p) \,D(P)\,\bar \chi_{\beta\delta}(q) \; +\;
\chi_{\gamma\alpha}^\mu(p) \,D^{\mu\nu}(P)
    \bar  \chi_{\beta\delta}^\nu(q) \nn \; ,   
\end{eqnarray}
$P$ is the total momentum of the incoming and the outgoing quark-quark
pair, $p$ and $q$ are the relative momenta between the quarks
in these channels as $y$ and $z$ are the relative coordinates.

$\chi_{\alpha\beta}(p)$ and $\chi_{\alpha\beta}^\mu(p)$ are 
vertex functions of quarks with a scalar and an axialvector
diquark, respectively.
They belong to a $\bar 3$-representation in color space and
are flavor antisymmetric (scalar diquark) or flavor
symmetric (axialvector diquark).
 For their Dirac structure we will retain
the dominant contribution only, and a scalar function $P(p)$ which depends
only on the relative momentum $p$ between the quarks parameterizes
the extension of the vertex in momentum space\footnote{
The Pauli principle requires then
the relative momentum to be defined $p=\frac{1}{2}(p_\alpha-p_\beta)$,
where $p_\alpha$ and $p_\beta$ are the quark momenta \cite{Oet99}.}:
\begin{eqnarray}
 \chi_{\alpha\beta}(p)&=&g_s (\gamma^5 C)_{\alpha\beta}\; P(p) \; ,
   \label{dqvertex_s} \\
 \chi_{\alpha\beta}^\mu(p)&=&g_a (\gamma^\mu C)_{\alpha\beta}\; P(p).
   \label{dqvertex_a}
\end{eqnarray}
$C$ denotes hereby the charge conjugation matrix and
$g_a$ and $g_s$ are normalization constants at this stage. The choice
\begin{equation}
P(p)=1
\end{equation}
corresponds to a point-like diquark whereas extended diquarks can be 
modeled as
\begin{equation}
P(p)= \left( \frac{\gamma^2}{\gamma^2+p^2} \right)^n.
\end{equation}
This specific form with $n$=2 or $n$=4 proved to be 
quite successful in describing
electromagnetic properties of the nucleon when using scalar diquarks
only \cite{Oet99}. 

To parameterize confinement, the propagators
 of scalar and axialvector diquark, appearing
in eq. (\ref{dq_sep}) as $D(P)$ and $D^{\mu\nu}(P)$, ought to be
modified. Our chosen form,
\begin{eqnarray}
D(p)&=& -\frac{1}{p^2+m_{sc}^2} \left( 1- e^{-\left(1+\frac{p^2}{m_{sc}^2}
   \right)} \right), \label{Ds}\\
D^{\mu\nu}(p)&=& -\frac{\delta^{\mu\nu}}{p^2+m_{ax}^2} 
  \left( 1- e^{-\left(1+\frac{p^2}{m_{ax}^2} \right)} \right), \label{Da} 
\end{eqnarray}
removes the free particle poles at the cost of an essential singularity
for time-like infinitely large momenta. 
The constituent quark propagator is modified likewise:
\begin{equation}
 S(p)= \frac{ i\pslash -m_q}{p^2+m_q^2} \left( 1- e^{-\left(1+\frac{p^2}{m_q^2}
   \right)} \right). \label{S}      
\end{equation}
With these ingredients, the Faddeev equations for the three quark system
can be simplified enormously.  
To do this, one enters the Faddeev equations with an ansatz for the
truncated, irreducible 3-quark correlation function 
(the 6-point quark Green function),
which, {\em e.g.}, exhibits a pole from a spin-1/2 baryon:
\begin{eqnarray}
  G_{\alpha\beta\gamma,\delta\epsilon\zeta}^{trunc}
 & \sim &
       \frac{  \Gamma_{\alpha\beta\gamma} (P;p,p_d,p_1) 
               \,\bar \Gamma_{\delta\epsilon\zeta} (P;q,q_d,q_1) }
            {P^2+M^2}, \\
   \Gamma_{\alpha\beta\gamma} &=& \chi_{\beta\gamma}(p_1) D(p_d)
       (\Phi^5(P,p) u)_\alpha
        + \chi^\mu_{\beta\gamma}(p_1)D^{\mu\nu}(p_d)(\Phi^\nu(P,p) u)_\alpha.
\end{eqnarray}     
The flavor and color indices which have to be found after projection onto
the baryon quantum numbers
have been omitted here. The object
of interest is now the nucleon vertex function
 $\Phi u=\pmatrix{\Phi \cr \Phi^\mu} u$
(with $u$ being a positive-energy Dirac spinor)
 which represents an effective spinor
characterizing the scalar and the axialvector diquark correlations
within the nucleon. 

\begin{figure}
\centerline{\epsfig{file=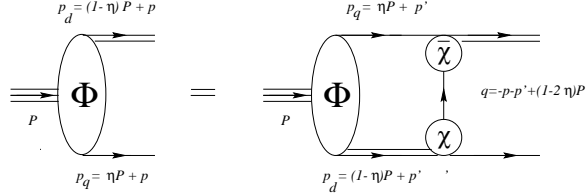, width=8cm}}
\caption{The baryon Bethe-Salpeter equation. The momentum partitioning
parameter $\eta$ distributes the relative momentum $p'$ over
quark and diquark.}
\label{bse}
\end{figure}

For this effective spinor, a coupled set of Bethe-Salpeter equations
 can be derived.  Its complete derivation 
can be found in \cite{Ish95}. For spin-1/2 baryons in the flavor-symmetric
case, the equation takes the form:
%
%
\begin{eqnarray}
 \pmatrix{\Psi^5 \cr \Psi^{\mu'}}(p,P) &:=& 
  S(p_q)  \pmatrix{ D & 0 \cr 0 & D^{\mu'\mu}} (p_d)
 \pmatrix{\Phi^5 \cr \Phi^\mu}(p,P) \label{bse_eq} \\
 \pmatrix{\Phi^5 \cr \Phi^\mu}(p,P)&=&
  \int \frac{d^4\,p'}{(2\pi)^4} \frac{1}{2}
  \pmatrix{-\chi S^T(q)\bar\chi & \sqrt{3} \chi^{\mu'} S^T(q)\bar\chi \cr
    \sqrt{3}\chi S^T(q)\bar\chi^{\mu} &  \chi^{\mu'} S^T(q)\bar\chi^\mu} 
  \pmatrix{\Psi^5 \cr \Psi^{\mu'}}(p',P).
   \nonumber 
\end{eqnarray}
It is pictorially represented in Fig. \ref{bse}. The attraction that leads
to a bound state is the quark exchange between the two constituents.
Note that 
we banned all unknown and possibly very complicated gluonic interactions 
between the quarks into the parameterization of the two-quark correlations.
The quark exchange is a consequence of the structure of the
Faddeev equations. The quark-diquark vertex from eqs. (\ref{dqvertex_s},
\ref{dqvertex_a})
enters as the quark-diquark interaction vertex.
This equation can be solved without any further approximation,
especially without any non-relativistic reduction. First one
decomposes the baryon vertex $\Phi$ (where each component is
a 4$\times$4-matrix) in Dirac space and projects onto positive
parity and energy states. This procedure is described in detail in
\cite{Oet98}. Choosing the rest frame of the nucleon, all independent
components are regrouped as eigenstates of orbital angular momentum.
As a final result, eight independent amplitudes, {\em i.e.}
scalar functions which multiply the components, describe the
spin-1/2 baryon as can be seen from Table \ref{vertex}.
\begin{table}[b]
\caption{Components of the octet baryon vertex function with their
respective spin and orbital angular momentum. 
$(\gamma_5 C)$ corresponds to
scalar and $(\gamma^\mu C),\,\mu=1 \dots 4,$ to axialvector
diquark correlations. Note that the partial waves in the first row
possess a non-relativistic limit.}
\begin{center}
\begin{tabular}{p{0.1cm}p{2.7cm}cccc}
 &&&&& \\
\multicolumn{2}{l}{\parbox{2.8cm}{ { \mbox{``non-relativistic''} partial waves} }  } &
  $\pmatrix{ \st \chi \cr \st 0 } \st{(\gamma_5 C)}$    &
 $\st\hat P^4{\pmatrix{ \st 0\cr \st \chi}} (\gamma^4 C)$    &
  ${\pmatrix{\st i\sigma^i\chi \cr \st 0}} \st (\gamma^i C)$    &
  ${\pmatrix{\st i\left(\hat p^i(\vec{\sigma}\hat{\vec{p}})-\frac{\sigma^i}{3}\right) \chi\cr \st 0}} \st (\gamma^i C)$ \\  
 &  {spin} & {1/2} & {1/2} & {1/2} & {3/2} \\
 &  {orbital angular momentum} & { $s$} & {$s$} & {$s$} & {$d$} \\
\multicolumn{2}{l}{\parbox{2.8cm}{ { ``relativistic'' \mbox{partial waves} }  }} &
  $\pmatrix{ \st 0 \cr \st \vec \sigma \vec p \chi } \st (\gamma_5 C)$    &
  $\st \hat P^4{\pmatrix{ \st (\vec{\sigma}\vec{p})\chi\cr \st 0}}(\gamma^4 C)$    &
  ${\pmatrix{\st 0\cr \st i\sigma^i(\vec{\sigma}\vec{p})\chi}}\st (\gamma^i C)$    &
  ${\pmatrix{\st 0\cr \st i\left(p^i-\frac{\sigma^i(\vec{\sigma}\vec{p})}{3}\right)\chi}} \st (\gamma^i C)$ \\   
 &  {spin} & \cl{1/2} & \cl{1/2} & \cl{1/2} & {3/2} \\
 & {orbital angular momentum} & \cl{ $p$} & \cl{$p$} & \cl{$p$} & {$p$}\\ \hline
\end{tabular}
\end{center}
\label{vertex}
\end{table}
As the amplitudes still depend on two momenta (the relative momentum
$p$ and the total momentum $P$), an expansion in terms of Chebyshev polynomials
for the variable $p\cdot P/(|p| |P|)$ is performed. Thus
the four-dimensional equation (\ref{bse_eq}) can be reduced to a number
of coupled one-dimensional integral equations \cite{Oet98,Oet99} which
we solved iteratively.

This procedure can be applied to spin-3/2 baryons as well\cite{Oet98}.
Again eight independent amplitudes are found after spin and
energy projection. Here, as a difference to spin-1/2 baryons, only
one $s$ partial wave exists which is found to dominate the expansion. 

\section{Results for Observables}
\subsection{Octet and Decuplet Masses}
In our approach the strange quark constituent mass
$m_s$ is the only source of flavour symmetry breaking. Isospin
is assumed to be conserved. The equations describing octet
and decuplet baryons have been derived under the
premises of flavour and spin conservation, {\em i.e.}
 only vertex function components
with same spin and flavour content couple. Again the full set
of equations can be found in \cite{Oet98}. The results for 
the cases of a pointlike diquark and an extended diquark are
shown in Tab. \ref{masses}. We chose scalar and axialvector diquark
masses\footnote{The use of confining propagators renders the masses 
to be mere parameters which set the scale in the propagators, eqs.
(\ref{Ds}-\ref{S}). They are of course unobservable.}
to be equal and proportional to the sum of the two quark masses
constituting the diquark. The proportionality constant is called $\xi$.
The nucleon and the delta mass served as input to determine
the normalization constants $g_s$ and $g_a$ appearing in eqs. 
(\ref{dqvertex_s},\ref{dqvertex_a}). From the viewpoint of the effective 
quark-diquark theory, $g_s$ and $g_a$ reflect the coupling strengths
in the two diquark channels. 
\begin{table}[t]
\caption{Octet and decuplet masses.}
\begin{center}
\begin{tabular}{llcc}
 & exp.  & pointlike & extended   \\
 &       & \multicolumn{2}{c}{diquark} \\
 &  & {${ P(p)=1}$} &
         {$ P(p)=\left(\frac{\ds\gamma^2}{\ds \gamma^2+p^2}\right)^4$}\\
  & & & { $\gamma=0.5$  GeV} \\ \hline
 &&& \\
$m_u\,$ {\small(GeV)}    &      &0.5       &0.56  \\
$m_s$ {\small(GeV)} &      &0.63    &0.68   \\
$\xi$     &      &0.73
&0.6  \\ \hline 
 &&& \\
$M_\Lambda\,$ {\small (GeV)} &1.116 &1.133& 1.098 \\
$M_\Sigma\,$ {\small (GeV)} &1.193  &1.140& 1.129 \\
$M_\Xi\,$ {\small (GeV)}     &1.315 &1.319& 1.279 \\ 
$M_{\Sigma^*}$ {\small (GeV)}&1.384 &1.380& 1.396\\
$M_{\Xi^*}$ {\small (GeV)}   &1.530 &1.516& 1.572\\
$M_\Omega\,$ {\small (GeV)}  &1.672 &1.665& 1.766\\ 
 &&&\\ \hline
\end{tabular}       
\end{center}
\label{masses}
\end{table}

As can be seen from the numbers, the mass splitting between octet
and decuplet can be explained as a result of the relativistic dynamics  
only. In the case of extended diquarks, the splitting is even
overestimated.

\subsection{Electromagnetic Form Factors}

Calculation of observables within the Bethe-Salpeter framework
proceeds along Mandelstam's formalism \cite{Man55}.
The two necessary ingredients are normalized nucleon-quark-diquark vertex
functions and, in case of the electromagnetic form factors, the current
operator. The vertex functions can be calculated as outlined  in the previous
section and their normalization is determined by the canonical
normalization to the correct (fermionic) bound state residue,
see, {\em e.g.}, \cite{Itz85}.
To this end, we define an object $G (p,p',P)$ involving the quark and diquark
 propagators and the
exchange kernel appearing in the Bethe-Salpeter equation (\ref{bse_eq}),
\begin{eqnarray}
 G (p,p',P) &=& (2\pi)^4 \delta(p-p') S^{-1}(p_q)
               \pmatrix{ D^{-1} & 0 \cr 0 & (D^{\mu'\mu})^{-1}} (p_d) +
   \nonumber \\
       & & \qquad \frac{1}{2}
  \pmatrix{\chi S^T(q)\bar\chi & -\sqrt{3} \chi^{\mu'} S^T(q)\bar\chi \cr
    -\sqrt{3}\chi S^T(q)\bar\chi^{\mu} & - \chi^{\mu'} S^T(q)\bar\chi^\mu}
   .
\end{eqnarray} 
With $\Lambda^+$ being the positive-energy projector, the normalization
condition is:
\begin{equation}
  -\int \frac{d^4\,p}{(2\pi)^4}
  \int \frac{d^4\,p'}{(2\pi)^4}
   \bar \Psi(p',P_n) \left[ P^\mu \frac{\partial}{\partial P^\mu}
    G (p,p',P) \right]_{P=P_n} \Psi(p,P_n) \stackrel{!}{=} M \Lambda^+.
\end{equation}
The current operator consists of the couplings of the photon
to quark and diquark (impulse approximation) and to the exchange kernel
$G$. For extended diquarks, it has been shown in \cite{Oet99} that
 the latter contribution encompasses 
two parts to make the total baryon current transversal
and to reproduce the correct charge. These two parts are the interaction
of the photon with the exchanged quark and its coupling to the
diquark-quark vertex $\chi$ or $\chi^\mu$ that can be described
by a seagull-like photon-quark-diquark vertex. In the case
of pointlike diquarks, this seagull contribution vanishes.

\begin{figure}
\caption{Diagrams that built up the baryon matrix elements of the
electromagnetic current. The first row shows the diagrams of
the impulse approximation, the second row the contributions
of the exchange kernel.}
\begin{center}
 \epsfig{file=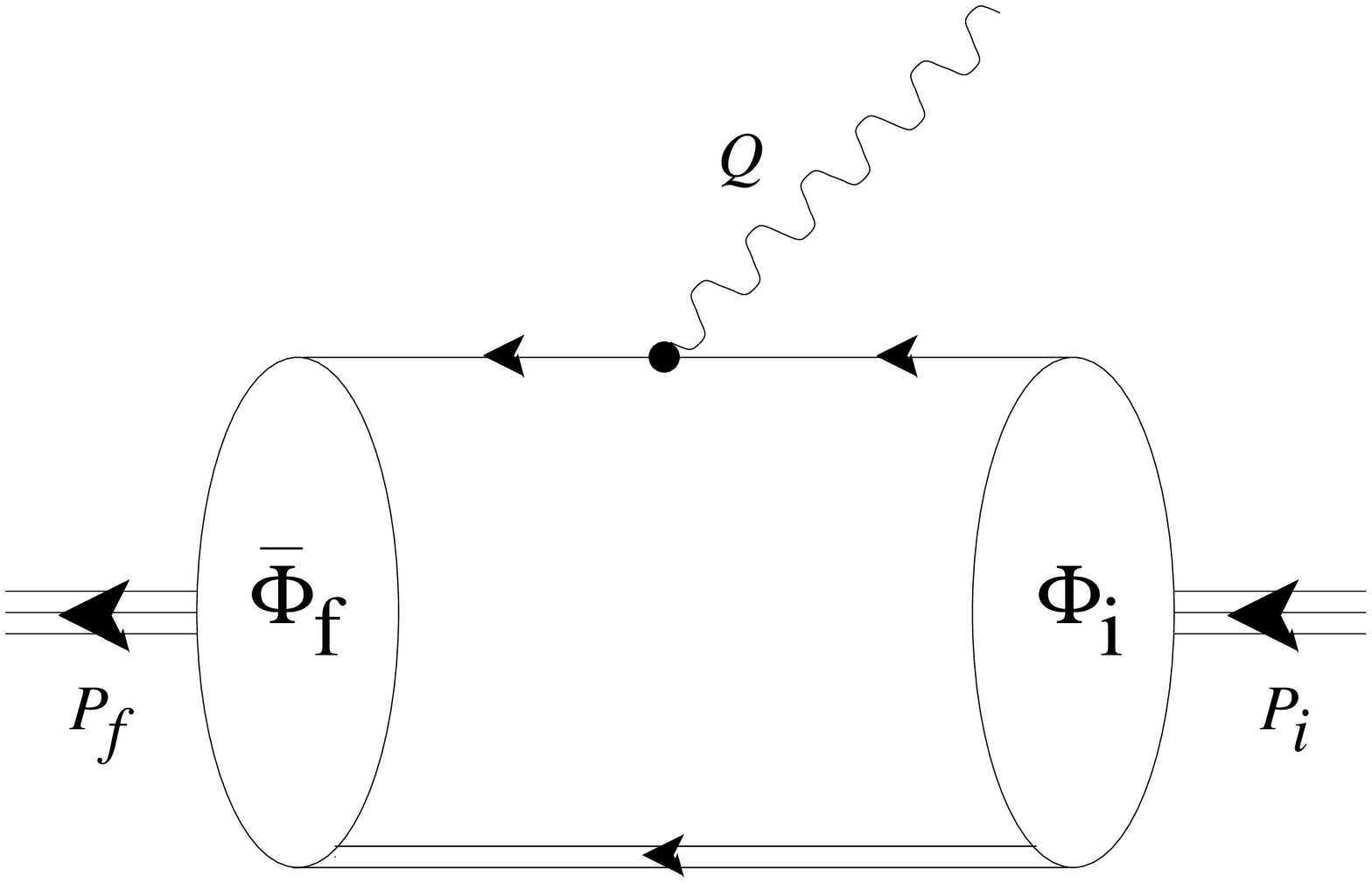,width=3.2cm}
 \begin{minipage}[t]{3.2cm}
  \epsfig{file=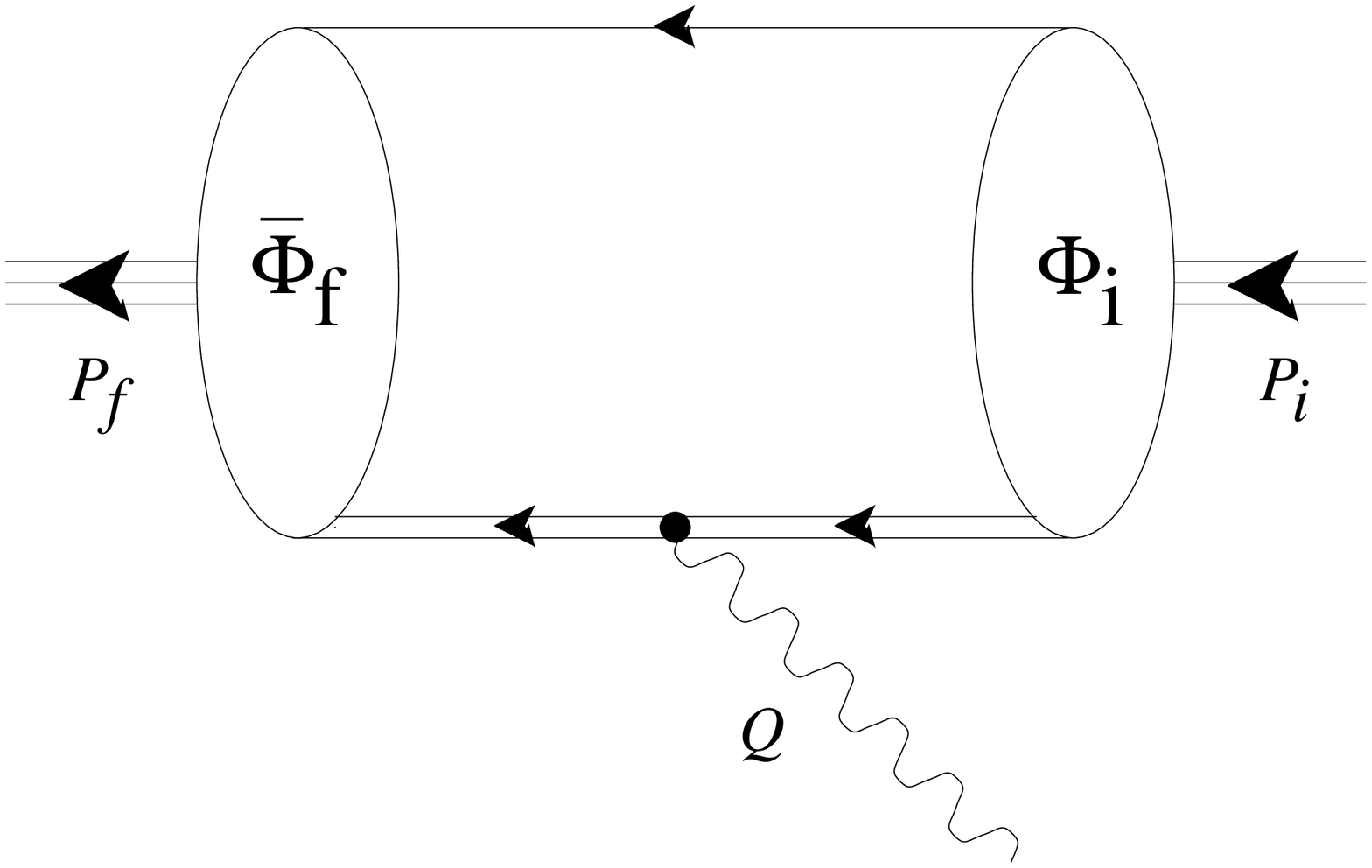,width=3.2cm}
 \end{minipage}
\end{center}
\begin{center}
 \epsfig{file=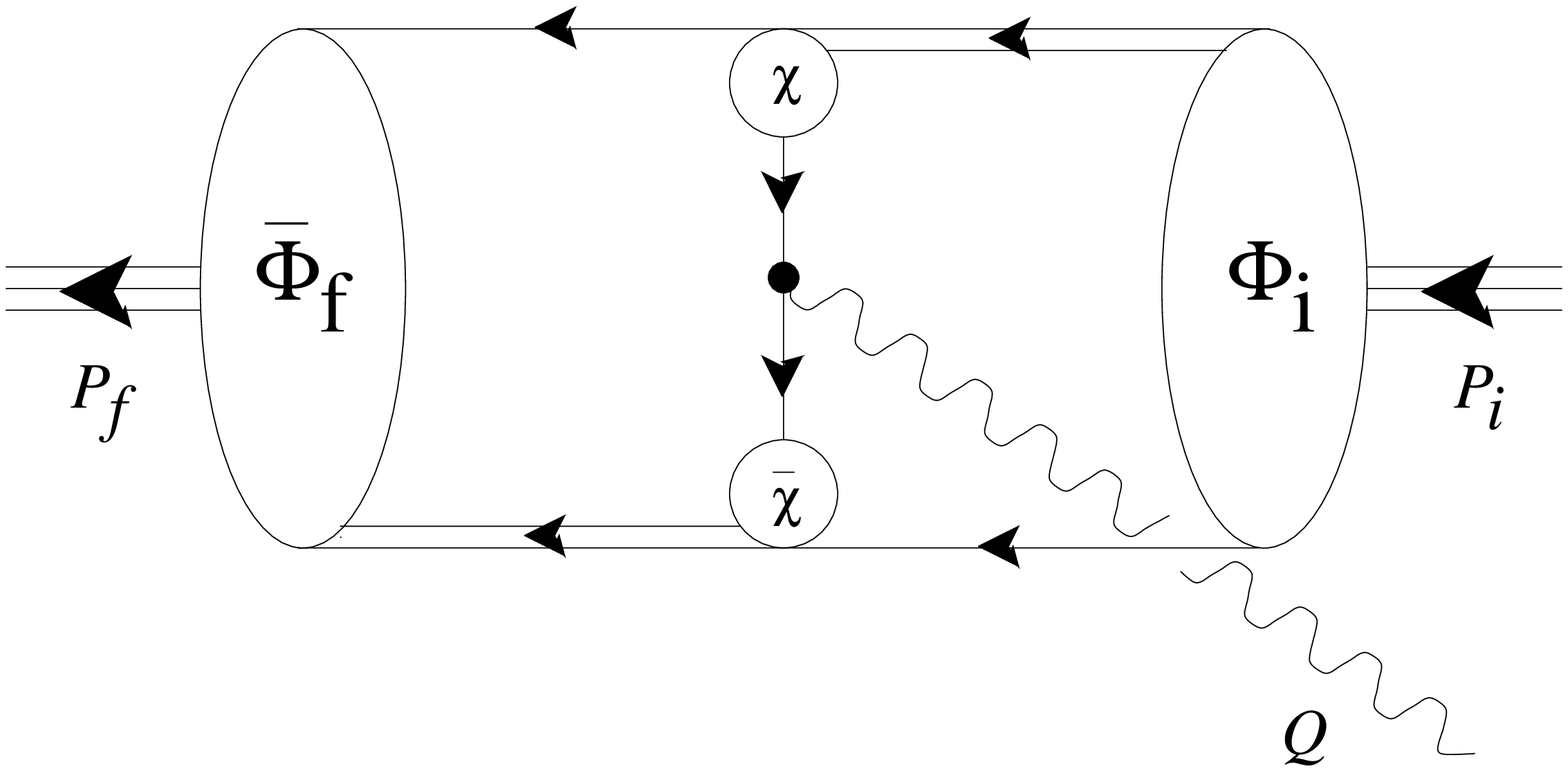,width=3.5cm}
 \epsfig{file=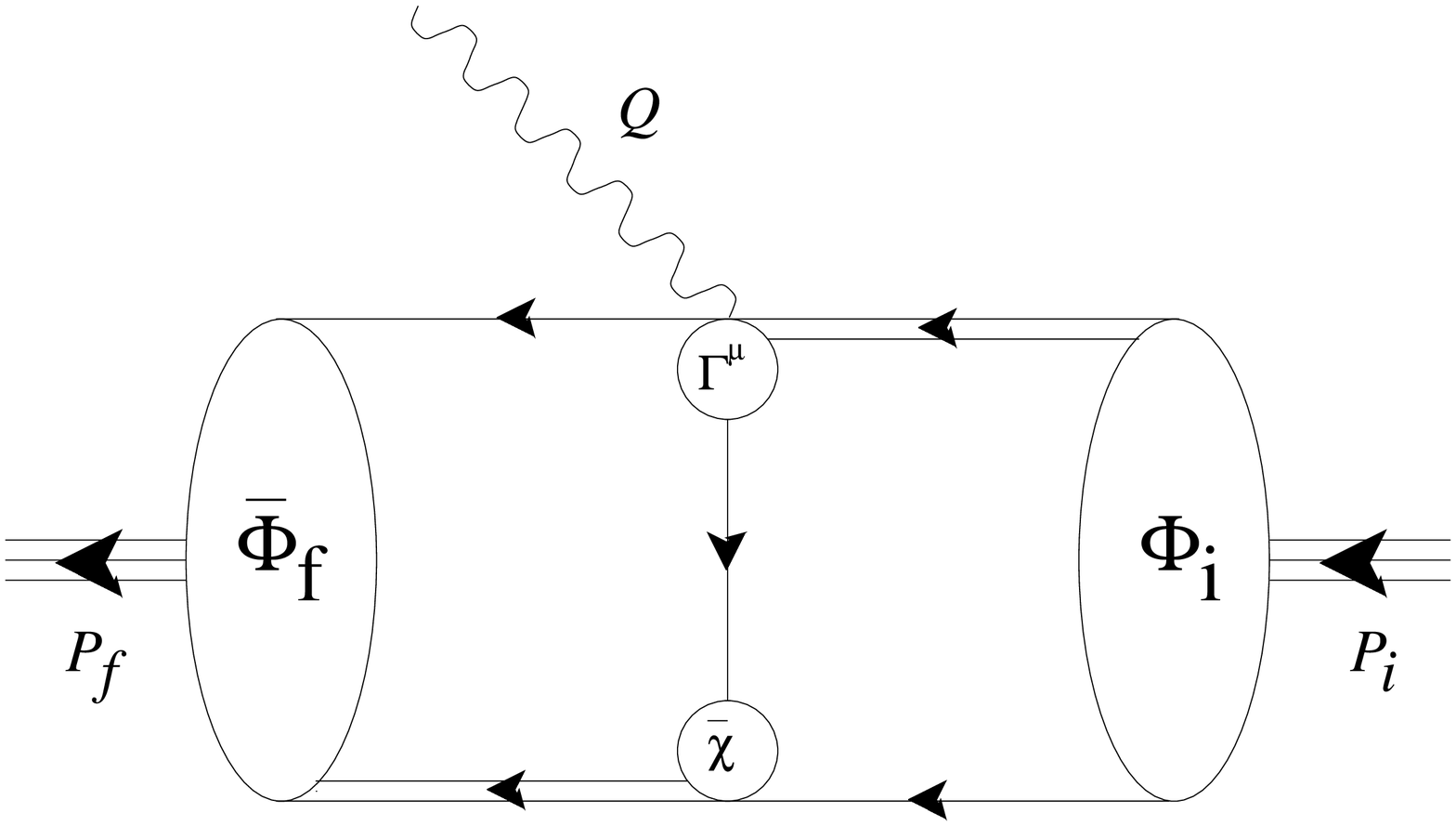,width=3.5cm}
 \epsfig{file=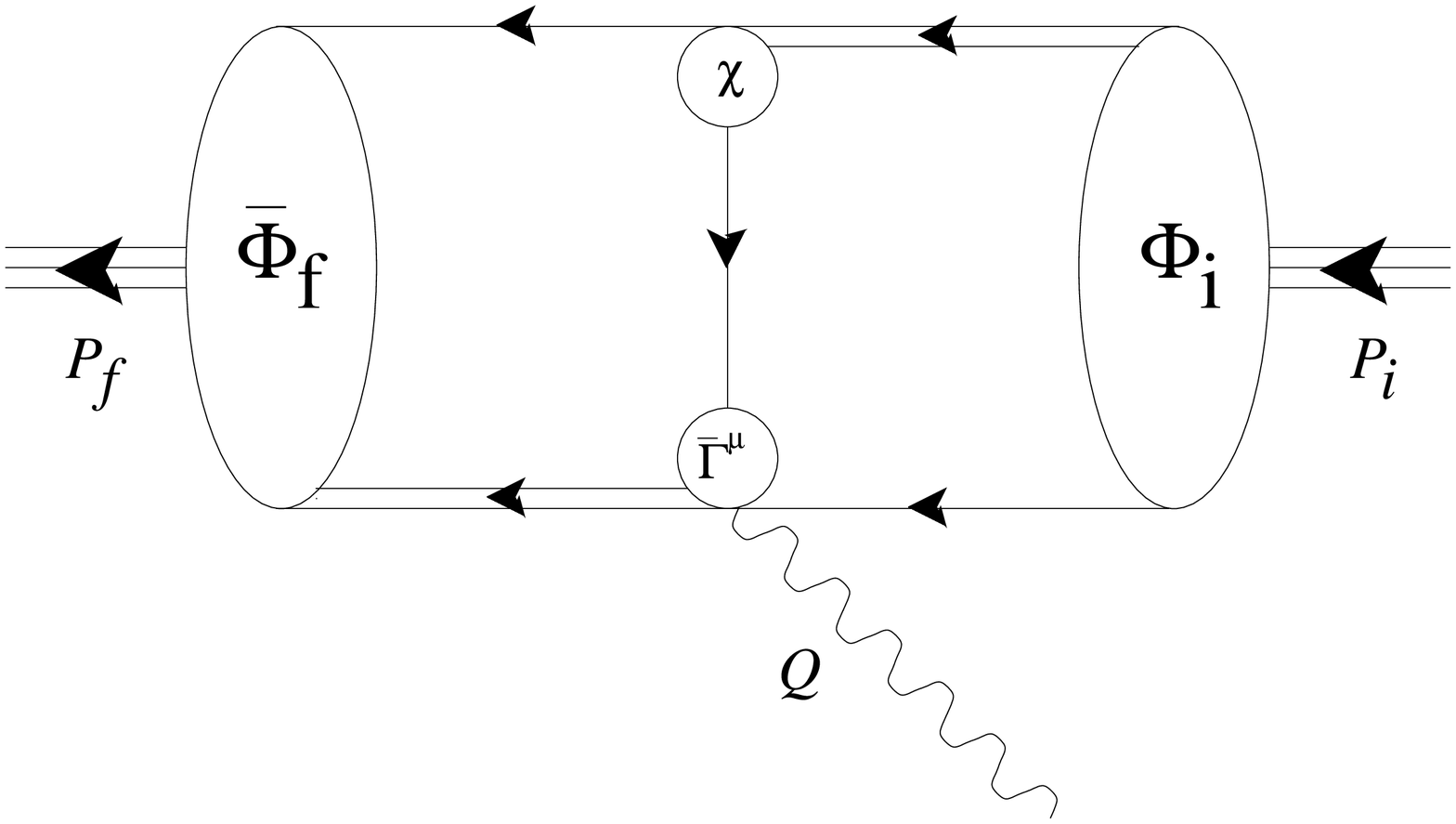,width=3.5cm}
\end{center}
\label{diags}
\end{figure}

To summarize, one has to calculate the diagrams given in Fig. \ref{diags}. 
To ensure gauge invariance, the quark-photon and the diquark-photon
vertices are of the Ball-Chiu type \cite{Bal80,Hel97a}. 
The seagull vertex is given by
\begin{equation}
  \Gamma^\mu=e_\alpha \frac{ 4p^\mu - Q^\mu}
{ 4p Q -Q^2} \,\left[\chi \left(p - \frac{Q}{2}\right)
-\chi (p)\right] -  \pmatrix{  \alpha \rightarrow \beta \cr
 Q \rightarrow -Q }.
\end{equation}
An analogous relation is valid for the seagull involving
the axialvector diquark vertex $\chi^\nu$. As before, $p$
is the relative momentum between the two quarks, and
$e_\alpha$, $e_\beta$ denote their respective charges. 

\begin{figure}[t]
\caption{The electric form factor of proton and neutron.}
\begin{center}
 \epsfig{file=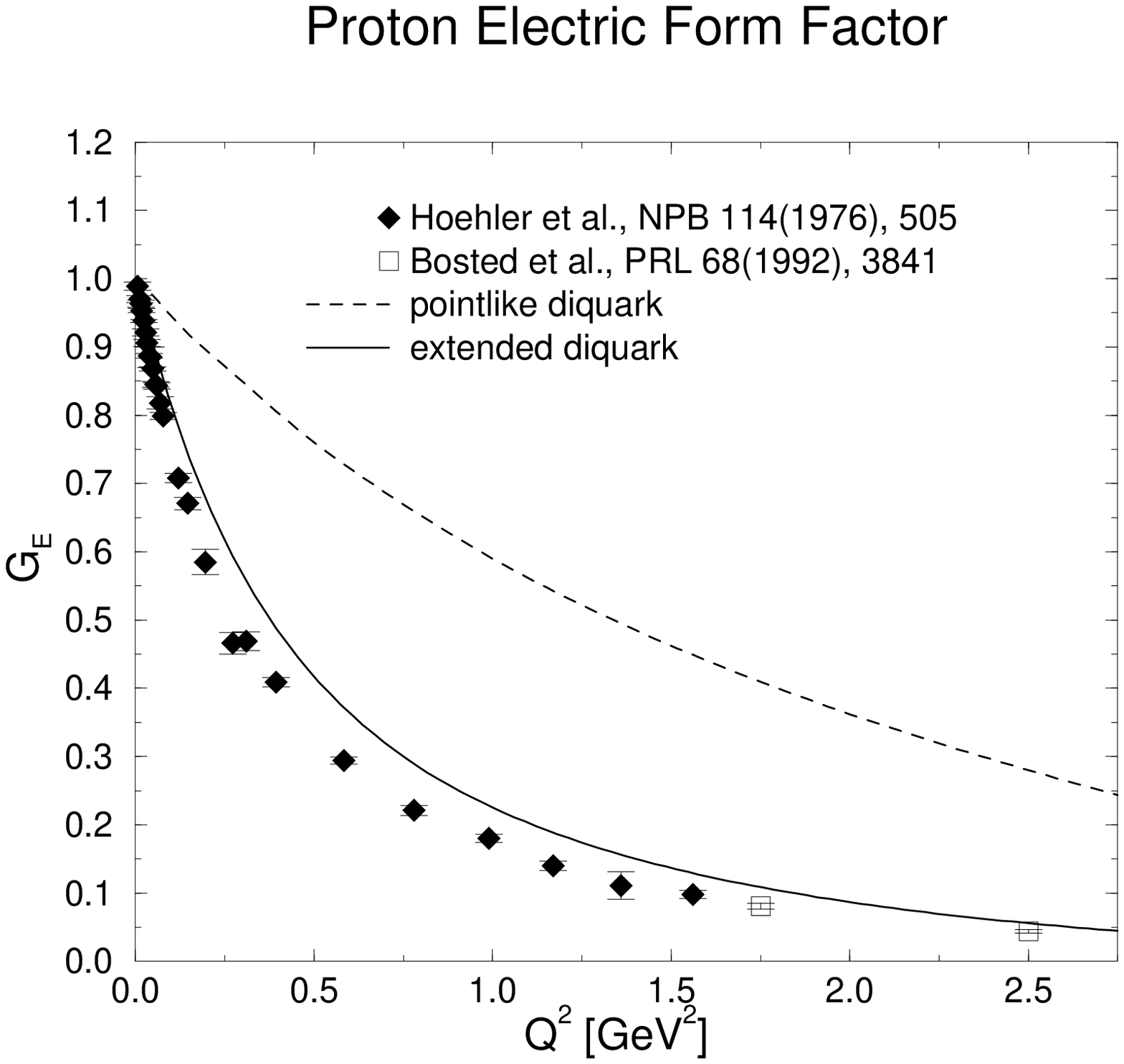,width=5cm}
 \epsfig{file=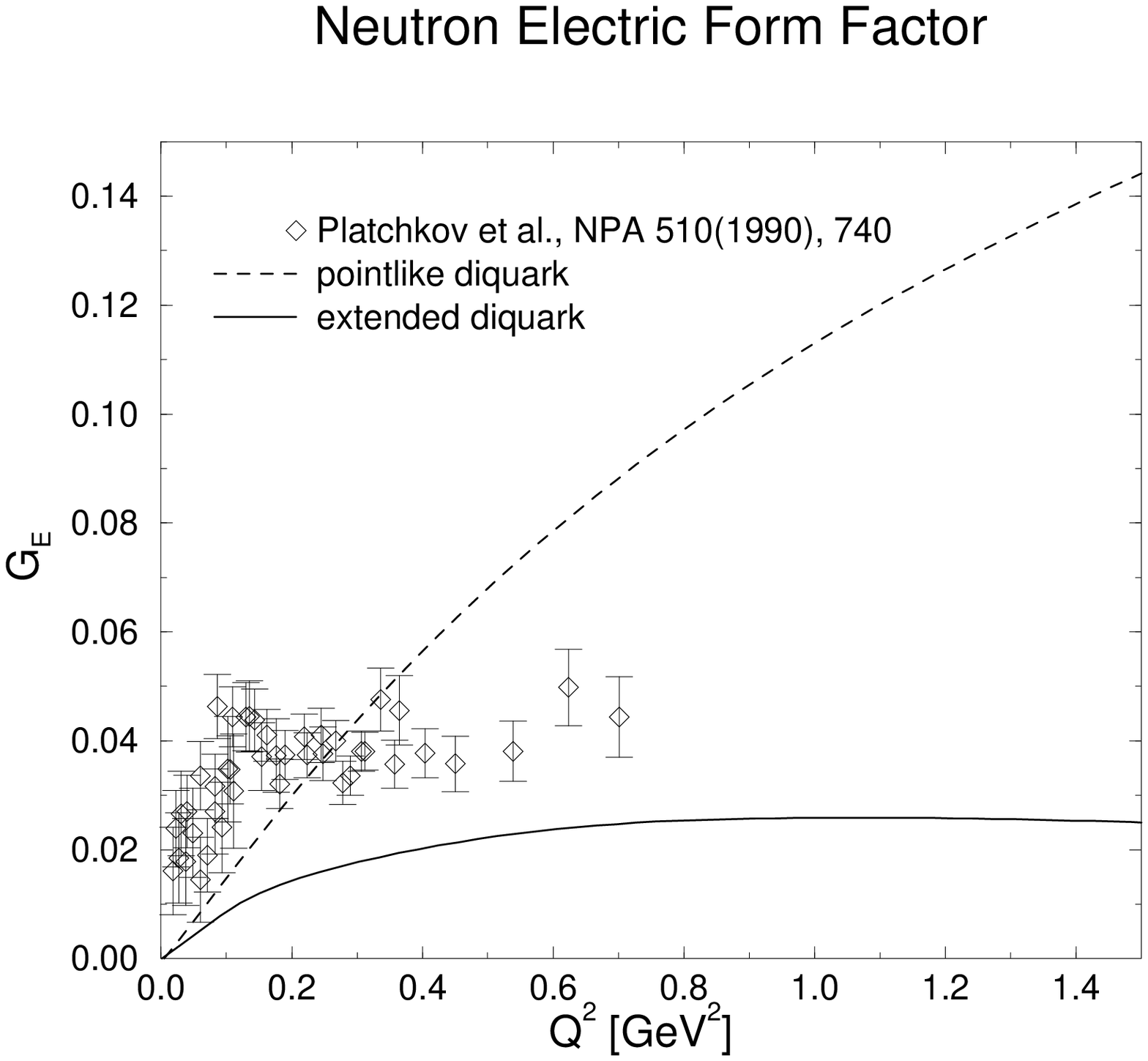,width=5cm}
\end{center}
\label{ge}
\end{figure}

We computed the Sachs form factors $G_E$ and $G_M$ for proton
and neutron using the parameters given in Tab. \ref{masses}.
The results for the electric form factor are shown in Fig. \ref{ge}.
Clearly, the proton curve falls too weakly for a pointlike diquark
which signals that the nucleon-quark-diquark vertex has too small
a size in coordinate space. This is remedied by the introduction
of the diquark structure. However, 
the neutron electric form  factor seems to be quenched too strongly
as compared to the data\footnote{As has been pointed out in \cite{Pas99},
these data should not be over-interpreted as systematic errors
have been involved in extracting them from raw data. Nevertheless they
give a feeling for the qualitative behaviour of the form factor.}.
Now this problem is probably due to overestimated axialvector diquark
correlations within the nucleon. Retaining extended scalar diquarks
only yields a very satisfactory description of the neutron $G_E$
\cite{Oet99}.

\begin{figure}[t]
\caption{The magnetic form factor of proton and neutron.}
\begin{center}
 \epsfig{file=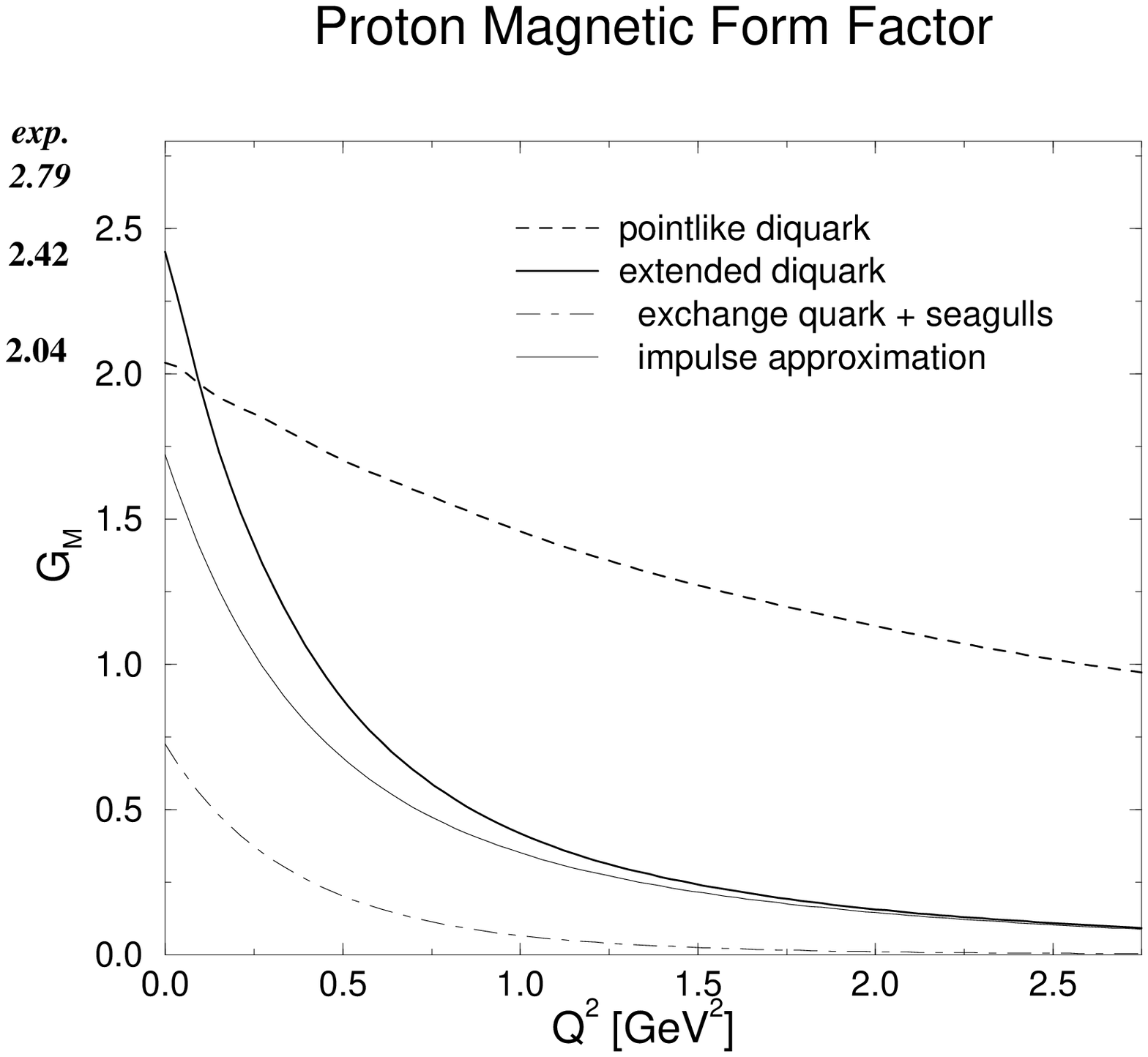,width=5cm}
 \epsfig{file=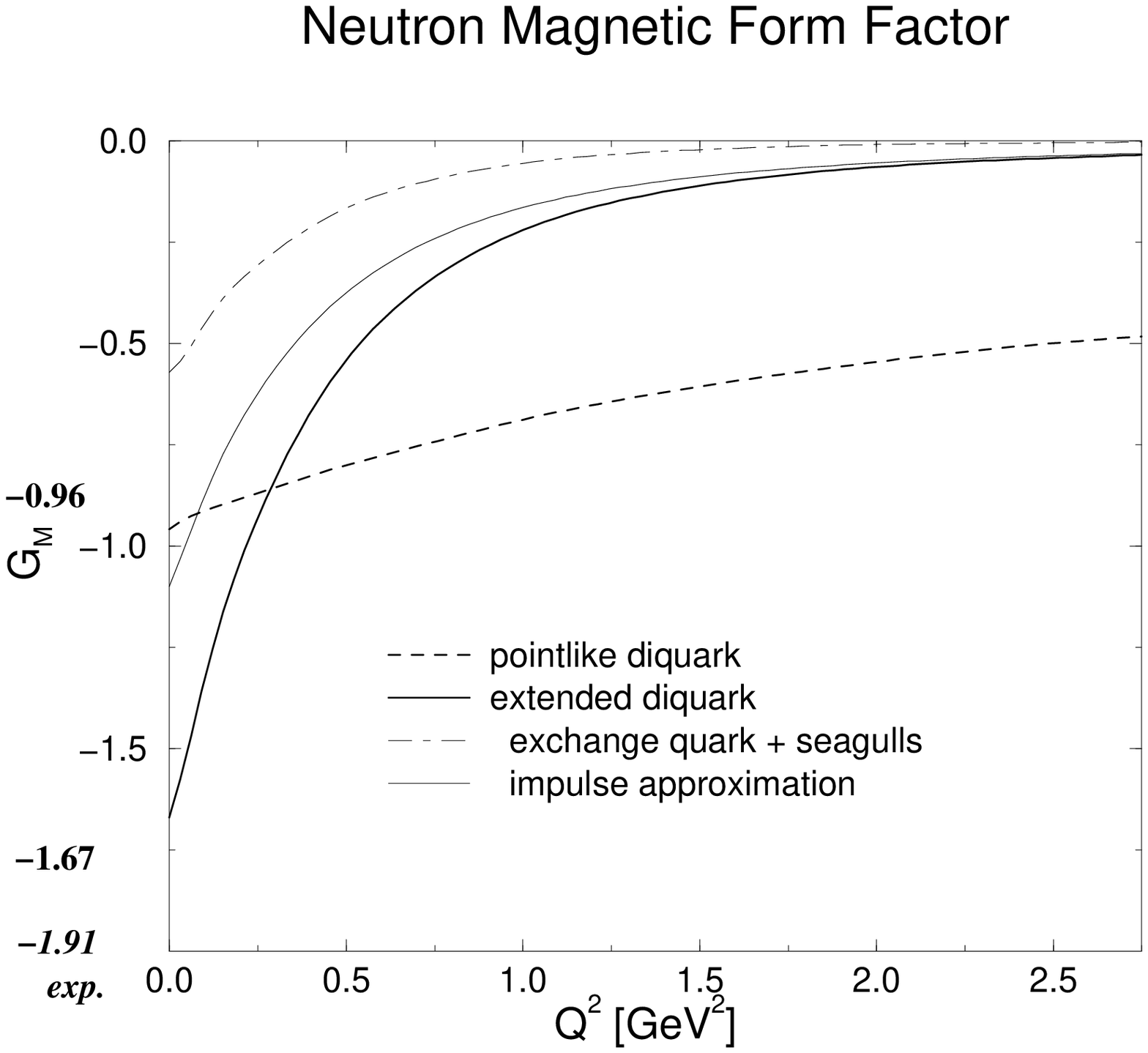,width=5cm}
\end{center}
\label{gm}
\end{figure}

The nucleon magnetic moments have also improved with the introduction
of the extended diquarks, see Fig. \ref{gm}. Nevertheless, their absolute
values are still about 13\% too small in comparison with experiment although
the ratio $\mu_p/\mu_n$ is reproduced nicely. In our formalism,
the diquarks have no anomalous magnetic moments since we do not
properly resolve the diquark in the second impulse approximation diagram
of Fig. \ref{diags}. Performing Mandelstam's formalism for the
diquark itself, {\em i.e.} coupling the photon to each of the quarks
and letting them recombine to the diquark, would therefore certainly
improve on the magnetic moments.  In Fig. \ref{gm} we have also plotted
separately the contributions of the impulse approximation and
of the coupling to the exchange kernel. As the second contribution
makes up more than 30 per cent of the total magnetic moment,
the less involved impulse approximation is merely a rough guide
to the behaviour of the magnetic form factor. 

\subsection{Strong Form Factors}

\begin{figure}[t]
\caption{The strong form factor $g_{\pi NN}$.}
\begin{center}
 \epsfig{file=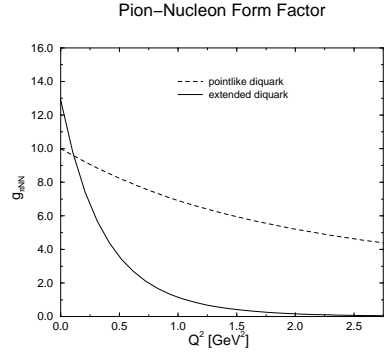,width=5cm}
\end{center}
\label{pion}
\end{figure}

Among various strong processes that are candidates for closer
scrutiny within our model, we have chosen first the pion-nucleon
form factor $g_{\pi NN}(Q^2)$. Hereby we couple the pion  to the quark only
with its dominant Dirac amplitude $\sim \gamma_5$. This is certainly
a good approximation as more detailed, microscopic calculations have shown
\cite{Tan97}. The on-shell pion-quark vertex is dictated by
PCAC and for the off-shell extrapolation we used a form proposed
by ref. \cite{Del79} and which has been applied in \cite{Hel97a}.
In our model, the diquark contributes nothing to $g_{\pi NN}$.
This is a simple consequence of the Dirac algebra if  one tries
to couple the pion to each of the two quarks within the diquark. 
The results for the impulse approximation diagram only is shown
in Fig. \ref{pion}. Again, the fall-off in the case of the pointlike
diquark is much slower than a monopole and appears to be unphysical.
In contrast to this, $g_{\pi NN}$ for the extended diquark
falls slightly stronger than a monopole with a width parameter
of around 360 MeV. In the light of the results for the magnetic moments,
the value of $g_{\pi NN}$ at $Q=0$ may still be subject to sizeable 
corrections coming from the coupling to the exchange quark.

\section{Conclusion}

We have suggested a field theoretic model of baryons that makes
use of diquarks which are a parameterization of
the quark-quark correlations within baryons. Thereby we could retain
full covariance. We parameterized confinement by a suitable modification
of quark and diquark propagators to avoid unphysical thresholds.

Masses and four-dimensional vertex functions have been calculated
for the baryon octet and decuplet. These vertex functions are
the main ingredient for the calculation of observables such as
the nucleon electromagnetic form factors. Whereas the mass spectrum is quite
unsensitive to the extension of the diquarks, the form factors
provide an effective mean to fix it. In these calculations
gauge invariance was strictly maintained. However, the nucleon magnetic 
moments are still about 15 per cent to small. This we attribute
to our incomplete handling of the electromagnetic structure
of the diquark. 

The computation of the pion-nucleon form factor is a necessary
intermediate step to calculate production processes.
As the pseudoscalar mesons do not couple to the diquarks,
these processes are particularly transparent
within the framework of our model.
Additionally, a $\Lambda$ hyperon in the final state
renders the flavor algebra simple, therefore we have chosen
associated strangeness production ($pp \rightarrow pK\Lambda$)
and kaon photoproduction ($\gamma p \rightarrow K\Lambda$)
as further testing ground for our model
\cite{Alk99}.\\[3mm] 
{\bf Acknowledgement:}  M.O. thanks the organizers for
the pleasant atmosphere at the workshop. The authors  also want
to express their gratitude to Hugo Reinhardt and Herbert Weigel 
for their support of this project.

\end{document}